\documentstyle[11pt]{article}
\begin{document}
\centerline{\LARGE Dissipative Systems and Objective Description:}\par
\centerline{\LARGE Quantum Brownian Motion as an Example}\par
\vskip 15  pt
\centerline{B.~Vacchini\footnote{Dipartimento di Fisica
dell'Universit\`a di Milano and Istituto Nazionale di Fisica
Nucleare, Sezione di Milano, Via Celoria 16, I-20133, Milan,
Italy. E-mail: bassano.vacchini@mi.infn.it}}
\vskip 15 pt
\centerline{\sc Abstract}\par
\vskip 15 pt
{
\baselineskip=12pt
A structure of generator of a quantum dynamical semigroup for the
dynamics of a test particle interacting through collisions with the
environment is considered, which has been obtained from a
microphysical model. The related master-equation is shown to go over
to a Fokker-Planck equation for the description of Brownian motion at
quantum level in the long wavelength limit. The structure of this
Fokker-Planck equation is expressed in this paper in terms of
superoperators, giving explicit expressions for the coefficient of
diffusion in momentum in correspondence with two cases of interest for
the interaction potential.  This Fokker-Planck equation gives an
example of a physically motivated generator of quantum dynamical
semigroup, which serves as a starting point for the theory of
measurement continuous in time, allowing for the introduction of
trajectories in quantum mechanics. This theory had in fact already
been applied to the problem of Brownian motion referring to similar
phenomenological structures obtained only on the basis of mathematical
requirements.
}\par
\pagenumbering{arabic}
\par
\section{INTRODUCTION}
\par
Despite its age the issue about the relationship between quantum and
classical world, perhaps most deeply stressed at the very beginning of
quantum mechanics by Niels Bohr, cannot be considered settled and
still gives rise to a lively debate, as confirmed for example by a
book recently published on the subject~\cite{Kiefer}, tackling it from
the point of view of decoherence. This very word has in fact recently
become very popular for the description of phenomena connected to the
transition from quantum to classical regime. While the interest in the
phenomenon of decoherence was previously mainly connected to
foundational issues, it is now mostly related to applications in
quantum computing. The extremely short time scales associated to the
phenomenon of decoherence, even in few-body systems, are in fact one
of the key problems to solve in order to leave the possibility open of
realizing in the future practically useful quantum
computers~\cite{ZeilingerQC}.

The term decoherence is used to denote the transition to dynamics
other than unitary even for few- or one-body systems, the effective
non-unitary subdynamics for these systems arising from the
impossibility to completely isolate them from the rest of the
laboratory, at least on sufficiently long time scales. As a result one
cannot expect that the physics of the microscopic system can be
correctly described by a unitary, reversible evolution driven by a
suitable self-adjoint Hamiltonian. Thus a simple picture in terms of a
Schr\"odinger equation fails, the correspondence principle is no
longer useful in order to envisage the generator of the dynamics, and
one is compelled to resort to a more general formalism.  In this
connection the studies on the foundations of quantum mechanics, in
particular on quantum structures~\cite{Grabowski} and on quantum
measurement theory~\cite{Busch}, have led to important results,
indicating possible new sceneries for quantum dynamics and especially
putting into evidence mathematical structures and properties relevant
for the quantum realm. More specifically a more modern formulation of
quantum mechanics has by now emerged~\cite{HolevoNEW}, where the
notions of effect (first introduced by
Ludwig~\cite{Foundations-Axiomatic}), coexistent observable,
POV-measure, operation and instrument allow for a better formulation
of irreversible dynamics and measurement processes. Based on these
concepts a formulation of continuous measurement theory in quantum
mechanics has been given, mainly developed by
Davies~\cite{Davies-continue}, the Milan
group~\cite{continue1-continue2-continue3-continue4-continue5} and
Holevo~\cite{Holevo-continue1-Holevo-continue2} (for an extensive
review see~\cite{HolevoNEW}). This theory relies on
the introduction of the generator of a quantum dynamical
semigroup~\cite{Gorini-qds-Alicki} for the dynamics of the observed
microscopic system, to which an operation-valued stochastic process
can be associated. It is then possible to introduce well-defined
functional probability densities in the space of time trajectories of
certain observables of the system, thus recovering, in this highly non
trivial way, elements of objective description, the very notion of
trajectory being a classical one (see~\cite{Lanz94} for a compact
review on the subject and~\cite{garda99-torun99-torun01} for a related approach to the
problem of objectivity in quantum mechanics). The observables for
which trajectories can be introduced depend on the very structure of
the quantum dynamical semigroup giving the irreversible time
evolution, the operators appearing in it and determining the
irreversible part of the dynamics also indicating the possible
measuring decompositions of the mapping giving the time evolution.

The general structure of bounded generators of quantum dynamical
semigroups, also satisfying the property of complete
positivity~\cite{Kraus,slovakia}, has been fully characterized by
Lindblad~\cite{Lindblad}, while in the unbounded case only a few
results are available~\cite{HolevoLNPH}. It is therefore of interest
to obtain physical examples of generators of quantum dynamical
semigroups, especially in the case in which the generator is
unbounded. In the following we will recall a result recently obtained
in this framework for the description of the motion of a test particle
in a quantum fluid~\cite{art1-art2,art3,reply,art5}, giving a new formulation in
terms of superoperators and further calculating the diffusion
coefficient for interaction potentials of physical interest. The
considered generator of quantum dynamical semigroups, obtained through
a microphysical derivation based on a scattering theory approach,
falls within a class known as quantum Brownian
motion~\cite{LindbladJMP}. This class of models has already been
considered within the framework of continuous measurement
theory~\cite{AlbertoQBM}, leading to a description in terms of
trajectories for the expectation values of the operators position and
momentum of the particle. The starting point for~\cite{AlbertoQBM} was
the phenomenological structure of generator of quantum Brownian motion
proposed by Lindblad~\cite{LindbladQBM,Sandulescu} on the basis of his
general result on completely positive quantum dynamical semigroups and
physical requirements on the dynamics originated from a classical
analogy. The result presented here gives a physically motivated
particular expression for the coefficients, determined in terms of
microphysical quantities, and for the selection of contributions
appearing in the structure of the generator.

\par
\section{MASTER EQUATION FOR A TEST PARTICLE IN A
QUANTUM GAS IN TERMS OF THE DYNAMIC \\ STRUCTURE FACTOR}
\setcounter{equation}{0}
\par

Let us consider the following problem of non-equilibrium statistical
mechanics: a test particle interacts through collisions with a fluid.
This model is known as Rayleigh gas~\cite{Spohn} and on a suitable time scale,
much longer than the typical relaxation time of the macroscopic fluid,
one expects a Markovian dynamics described in terms of a
master-equation. In the quantum case an expression has recently been
proposed for the generator of such a dynamics, which is in particular
the generator of a completely positive
quantum dynamical semigroup~\cite{art5}. The master-equation takes the 
following form
\begin{equation}
  \label{eq}
        {  
        d {\hat \varrho}  
        \over  
                      dt
        }  
        =
        -
        {i \over \hbar}
        [{\hat {\sf H}}_0
        ,
        {\hat \varrho}
        ]
        +
        {\cal L} [\hat \varrho],
\end{equation}
where $\hat \varrho$ is the statistical operator associated to the
test particle of mass $M$, ${\hat {\sf H}}_0$ the free Hamiltonian
${\hat {\sf p}}^2 / 2M$ and the mapping giving the dissipative part of 
the time evolution has the following Lindblad structure
\begin{equation}
  \label{l}
        {\cal L}[\cdot]= {2\pi \over\hbar}
        (2\pi\hbar)^3
        n
        \int_{{\bf R}^3} d^3\!
        {\bf{q}} \,  | \tilde{t} (q) |^2
        \Biggl[
        \hat{U} ({\bf{q}})
        \sqrt{
        S({\bf{q}},{\hat {\sf{p}}})
        }
        \cdot
        \sqrt{
        S({\bf{q}},{\hat {\sf{p}}})
        }
        \hat{U}^{\dagger} ({\bf{q}})
        -
        \frac 12
        {
        \left \{
        S({\bf{q}},{\hat {\sf{p}}}),
        \cdot
        \right \}
        }
        \Biggr].
\end{equation}
The unitary operators $\hat{U} ({\bf{q}})$ are given by
$e^{{i\over\hbar}{\bf{q}}\cdot{\hat 
    {\sf{x}}}}$, while the function $ \tilde{t} (q)$ is the Fourier
transform with respect to the transferred momentum ${\bf{q}}$ of the 
T-matrix describing the collisions between test particle and fluid,
supposed to depend only on the modulus of the momentum transfer and
in a negligible way on energy. The function $S({\bf{q}},
{\bf{p}})$
appearing operator-valued in (\ref{l}) is a positive two-point
correlation function known in the physical community as
dynamic structure factor~\cite{Lovesey,Griffin}, and it is usually
expressed as a function of momentum and energy transfer, ${\bf{q}}$
and $E$. It is defined by
\begin{equation}
  \label{dsf}
  S({\bf{q}},E)\equiv S({\bf{q}},{\bf{p}}) = 
        {  
        1  
        \over  
         2\pi\hbar
        }  
        \int_{{\bf R}} dt 
        {\int_{{\bf R}^3} d^3 \! {\bf{x}} \,}        
        e^{
        {
        i
        \over
         \hbar
        }
        [E ({\bf{q}},{\bf{p}}) t -
        {\bf{q}}\cdot{\bf{x}}]
        } 
      \frac{1}{N}
        {\int_{{\bf R}^3} d^3 \! {\bf{y}} \,}
        \left \langle  
         N({\bf{y}})  
         N({\bf{y}}+{\bf{x}},t)  
        \right \rangle        ,
\end{equation}
with 
\begin{displaymath}
    E ({\bf{q}},{\bf{p}})\equiv E=
{
({\bf{p}}+{\bf{q}})^2
\over
   2M
}
-
{
p^2
\over
   2M
}
=
{
q^2
\over
   2M
}
+
{
        {\bf{p}}
        \cdot
        {\bf{q}}
\over
M
}
\end{displaymath}
thus being the Fourier transform of the two-point time dependent
density correlation function of the fluid, calculated with respect to
the statistical operator describing the fluid at equilibrium. The
dynamic structure factor is always positive since it is proportional to 
the energy dependent scattering cross-section of a microscopic probe
off a macroscopic sample~\cite{vanHove}, and it gives the spectrum of
spontaneous fluctuations of the macroscopic sample.

In particular the dynamic structure factor can be exactly calculated in 
the case of a free quantum gas, thus obtaining close expressions for
Bose-Einstein and Fermi-Dirac statistics, which both go over to
Maxwell-Boltzmann statistics in the limit of low density. Denoting by 
$S_{\rm \scriptscriptstyle BE}({\bf{q}},{\bf{p}})$ the
dynamic structure factor for a free gas of particles of mass $m$
obeying Bose-Einstein statistics one has~\cite{art5}
\begin{eqnarray}
  \label{BE}
        S_{\rm \scriptscriptstyle BE}({\bf{q}},{\bf{p}})
        &=&{}
        -
        {
        1
        \over
         (2\pi\hbar)^3
        }
        {
        2\pi m^2
        \over
        n\beta q
        }
        {
        1
        \over
        1-
        \exp{
        \left[  
        {
        \beta
        \over
             2m
        }
        \left(
        2\sigma({\bf{q}},{\bf{p}})q -q^2
        \right)
        \right]
        }
        }
      \\
      &&\!\!\!\!\times\nonumber
        \log
        \left[
          1
          -
          \left\{1-        \exp{
        \left[  
        {
        \beta
        \over
             2m
        }
        \left(
        2\sigma({\bf{q}},{\bf{p}})q -q^2
        \right)
        \right]
        }
      \right\}
        {
        \exp{
        \left[  
        -{
        \beta
        \over
             2m
        }
        \sigma^2({\bf{q}},{\bf{p}})
        \right]
        }
        \over
        1- z
        \exp{
        \left[  
        -{
        \beta
        \over
             2m
        }
        (\sigma({\bf{q}},{\bf{p}}) - q)^2
        \right]
        }
        }
        \right]  
\end{eqnarray}
with $\beta=1/kT$ the inverse of the temperature, $n$ the particle
density, $z$ the fugacity of the gas, which is a number positive and
less than one for Bose-Einstein particles~\cite{Pathria}, and
\begin{displaymath}
    \sigma({\bf{q}},{\bf{p}})=
\frac{1}{2q}
\left[
q^2+2m E({\bf{q}},{\bf{p}})
\right].
\end{displaymath}
Similarly for Fermi-Dirac statistics
\begin{eqnarray}
  \label{FD}
        S_{\rm \scriptscriptstyle FD}({\bf{q}},{\bf{p}})
        &=&{}
        +
        {
        1
        \over
         (2\pi\hbar)^3
        }
        {
        2\pi m^2
        \over
        n\beta q
        }
        {
        1
        \over
        1-
        \exp{
        \left[  
        {
        \beta
        \over
             2m
        }
        \left(
        2\sigma({\bf{q}},{\bf{p}})q -q^2
        \right)
        \right]
        }
        }
      \\
      &&\!\!\!\!\times\nonumber
        \log
        \left[
          1
          +
          \left\{1-        \exp{
        \left[  
        {
        \beta
        \over
             2m
        }
        \left(
        2\sigma({\bf{q}},{\bf{p}})q -q^2
        \right)
        \right]
        }
      \right\}
        {
        \exp{
        \left[  
        -{
        \beta
        \over
             2m
        }
        \sigma^2({\bf{q}},{\bf{p}})
        \right]
        }
        \over
        1+ z
        \exp{
        \left[  
        -{
        \beta
        \over
             2m
        }
        (\sigma({\bf{q}},{\bf{p}}) - q)^2
        \right]
        }
        }
        \right],
\end{eqnarray}
so that the difference only lies in a suitable change of signs and in
the range of the fugacity $z$ which is positive without further
restrictions for Fermi-Dirac particles~\cite{Pathria}. Both (\ref{BE}) 
and (\ref{FD}) in the limit of low density, corresponding to $z$ much
smaller than one, 
lead in a straightforward way to the expression for a gas of
Maxwell-Boltzmann particles, as can be seen expanding the logarithm:
\begin{equation}
        \label{MB}
        S_{\rm \scriptscriptstyle MB}({\bf{q}},{\bf{p}})
        =
        {
        1
        \over
         (2\pi\hbar)^3
        }
        {
        2\pi m^2
        \over
        n\beta q
        }
        z
        \exp{
        \left[  
        -{
        \beta
        \over
             2m
        }
        \sigma^2({\bf{q}},{\bf{p}})
        \right]
        },
        \end{equation}
where the fugacity is now given by the explicit expression
\begin{displaymath}
  z=n\left(\frac{2\pi \hbar^2\beta}{m} \right)^{3/2}.
\end{displaymath}

A case of particular interest in which to apply (\ref{eq}) is the
description at quantum level of Brownian motion, that is the case in
which the mass $M$ of the test particle is much bigger than the mass
$m$ of the gas particles. One therefore needs expressions for the
dynamic structure factor in the Brownian limit in which the ratio
$\alpha=m/M$ is much smaller than one. To do this one writes the
argument of the exponentials in (\ref{BE}), (\ref{FD}) and (\ref{MB})
as a polynomial in $\alpha$, keeping only the contributions in the
lowest order. Concentrating on the simplest case of a gas of
Maxwell-Boltzmann particles, writing $\sigma^2$ as a polynomial in
$\alpha$
\begin{displaymath}
  \sigma^2({\bf{q}},{\bf{p}})
  =
  \frac{q^2}{4}
  + \frac{1}{2}\alpha[q^2 +2 {\bf{p}}
  \cdot {\bf{q}}]
  + \frac{1}{4}\frac{\alpha^2}{q^2}[q^2 +2 {\bf{p}}
  \cdot {\bf{q}}]^2,
\end{displaymath}
and keeping terms up to first order one has
\begin{equation}
        \label{8}
        S^{\scriptscriptstyle\infty}_{\rm \scriptscriptstyle
          MB}({\bf{q}},{\bf{p}}) 
        =
        {
        1
        \over
         (2\pi\hbar)^3
        }
        {
        2\pi m^2
        \over
        n\beta q
        }
        z
        e^{
        -{
        \beta
        \over
             8m
        }
        q^2
        }
        e^{
        -\frac{\beta}{2}
        [
        {
        q^2
        \over
             2M
        }
        +
        {{\bf{q}}\cdot{\bf{p}} \over M}
        ]
        },
        \end{equation}
where the index $\infty$ denotes the Brownian limit $\alpha\ll
1$. Eq.~(\ref{eq}) now becomes
\begin{eqnarray}
  \label{100}
        {  
        d {\hat \varrho}  
        \over  
                      dt
        }  
        &=&
        -
        {i \over \hbar}
        [
        {\hat {{\sf H}}}_0
        ,
        {\hat \varrho}
        ]
        +
        {2\pi \over\hbar}
        (2\pi\hbar)^3
        n
        \int_{{\bf R}^3} d^3\!
        {{\bf q}}
        \,  
        {
        | \tilde{t} (q) |^2
        }
      \nonumber
      \\
        &&
        \hphantom{cosicosi}
        \times
      \Biggl[\hat{U}^{\dagger} ({\bf{q}})
                \sqrt{
        S^{\scriptscriptstyle\infty}_{\rm \scriptscriptstyle MB}({{\bf
            q}},{\hat {{\sf p}}}) 
        }
        {\hat \varrho}
        \sqrt{
        S^{\scriptscriptstyle\infty}_{\rm \scriptscriptstyle MB}({{\bf
            q}},{\hat {{\sf p}}}) 
        }
         \hat{U} ({\bf{q}})-
        \frac 12
        \left \{
        S^{\scriptscriptstyle\infty}_{\rm \scriptscriptstyle MB}({{\bf
            q}},{\hat {{\sf p}}}), 
        {\hat \varrho}
        \right \}
        \Biggr]
\end{eqnarray}
and in view of (\ref{8}), introducing the operators
\begin{displaymath}
          V({\bf{q}},{\hat {{\sf p}}},{\hat
        {{\sf x}}})
        =
        e^{{i\over\hbar}{\bf{q}}\cdot{\hat {{\sf x}}}}
        e^{-{\beta\over 4M}{\bf{q}}\cdot{\hat {{\sf p}}}}
\end{displaymath}
(\ref{100}) takes the more manifest Lindblad structure
\begin{eqnarray}
        \label{riscoperta}
        \nonumber
        {  
        d {\hat \varrho}  
        \over  
                      dt
        }  
        &=&
        {}-
        {i \over \hbar}
        \left[{\hat {{\sf H}}}_0
        ,
        {\hat \varrho}
        \right]
        +
        z{4\pi^2 m^2 \over\beta\hbar}
        \int_{{\bf R}^3} d^3\!
        {{\bf q}}
        \,  
        {
        | \tilde{t} (q) |^2
        \over
        q
        }
        e^{-
        {
        \beta
        \over
             8m
        }
        {(1+2\alpha) {{q}}^2}
        }
        \\
        &&
        \hphantom{cosicosi}
        \times
        \Biggl[V({\bf{q}},{\hat {{\sf p}}},{\hat
        {{\sf x}}})
        {\hat \varrho}
        V^\dagger({\bf{q}},{\hat {{\sf p}}},{\hat
        {{\sf x}}})
        -
        \frac 12
        \left \{
        V^\dagger({\bf{q}},{\hat {{\sf p}}},{\hat
        {{\sf x}}})
        V({\bf{q}},{\hat {{\sf p}}},{\hat
        {{\sf x}}})
      ,
      {\hat \varrho}
        \right \}
        \Biggr]
        \nonumber
        \\
        &=&
        {}-
        {i \over \hbar}
        \left[ {\hat {{\sf H}}}_0       ,
        {\hat \varrho}
        \right]
        +
        z{4\pi^2 m^2 \over\beta\hbar}
        \int_{{\bf R}^3} d^3\!
        {{\bf q}}
        \,  
        {
        | \tilde{t} (q) |^2
        \over
        q
        }
        e^{-
        {
        \beta
        \over
             8m
        }
        {(1+2\alpha) {{q}}^2}
        }
      \nonumber
        \\
        &&
        \hphantom{cosicosi}
        \times
        \Biggl[
        e^{{i\over\hbar}{{\bf q}}\cdot{\hat {{\sf x}}}}
        e^{-{\beta\over 4M}{{\bf q}}\cdot{\hat {{\sf p}}}}
        {\hat \varrho}
        e^{-{\beta\over 4M}{{\bf q}}\cdot{\hat {{\sf p}}}}
        e^{-{i\over\hbar}{{\bf q}}\cdot{\hat {{\sf x}}}}
        - {1\over 2}
        \left \{
        e^{-{\beta\over 2M}{{\bf q}}\cdot{\hat {{\sf p}}}}
        ,
        {\hat \varrho}
        \right \}
        \Biggr].
\end{eqnarray}
The action of the operators position and momentum of the microsystem
${\hat {{\sf x}}}$ and
${\hat {{\sf p}}}$ is best seen introducing the following
superoperators
\begin{eqnarray}
  \label{super}
  {\cal L}^{\scriptscriptstyle -}_{\scriptscriptstyle\hat{\sf
      A}}[\cdot]&=&\frac{i}{\hbar}è[\hat{\sf
    A},\cdot]_{-}=\frac{i}{\hbar}è[\hat{\sf A},\cdot] 
  \\ \nonumber
  {\cal L}^{\scriptscriptstyle +}_{\scriptscriptstyle\hat{\sf
      A}}[\cdot]&=&\frac{1}{\hbar}è[\hat{\sf
    A},\cdot]_{+}=\frac{1}{\hbar}è\{\hat{\sf A},\cdot\},
\end{eqnarray}
which will also prove useful for future expansions. In terms of
(\ref{super}) eq.~(\ref{riscoperta}) takes the remarkably simple structure
\begin{eqnarray}
        \label{superprl}
        {  
        d {\hat \varrho}  
        \over  
                      dt
        }  
        &=&
        {}-
        {i \over \hbar}
        \left[{\hat {{\sf H}}}_0
        ,
        {\hat \varrho}
        \right]
        +
        z{4\pi^2 m^2 \over\beta\hbar}
        \int_{{\bf R}^3} d^3\!
        {{\bf q}}
        \,  
        {
        | \tilde{t} (q) |^2
        \over
        q
        }
        e^{-
        {
        \beta
        \over
             8m
        }
        {(1+2\alpha) {{q}}^2}
        }
        \\
        &&
        \hphantom{cosicosi}
        \times
        \Biggl[
        \exp\left({{\cal L}^{\scriptscriptstyle -}_{\scriptscriptstyle
              {{\bf q}}\cdot{\hat {{\sf x}}}}}\right) 
        \exp\left({{\cal L}^{\scriptscriptstyle +}_{\scriptscriptstyle
              \kappa{{\bf q}}\cdot{\hat {{\sf p}}}}}\right) 
        [{\hat \varrho}]
        -
        \frac 12
        \left \{
      \exp\left({\frac{2\kappa}{\hbar}{{\bf q}}\cdot{\hat {{\sf p}}}}\right)
          ,
      {\hat \varrho}
        \right \}
        \Biggr]
        \nonumber
\end{eqnarray}
with $\kappa=-\frac{\beta\hbar}{4M}$. The master-equation
(\ref{riscoperta}) gives a physical realization of a general structure
of generators of translation-covariant quantum dynamical semigroups
recently introduced by Holevo~\cite{HolevoTI}. In fact
(\ref{riscoperta}) and more generally (\ref{eq}) are invariant under
spatial translations in the sense that
\begin{equation}
  \label{trasla}
  {\cal L}[{\cal U}_{\bf a}[\hat{\sf w}]]={\cal U}_{\bf a}[{\cal
    L}[\hat{\sf w}]], 
\end{equation}
with $\hat{\sf w}$ a statistical operator and ${\cal U}_{\bf a}
[\cdot]= e^{-{i\over\hbar}{\bf{a}}\cdot{\hat {\sf{p}}}}\cdot
e^{+{i\over\hbar}{\bf{a}}\cdot{\hat {\sf{p}}}}$. In particular,
provided the macroscopic system is in a $\beta$-KMS state~\cite{Haag},
thus implying the detailed balance condition for the dynamic structure
factor~\cite{Brenig}, a stationary solution of (\ref{eq}) is given by
\begin{equation}
  \label{equi}
  \hat{\sf w}_0 (\hat{\sf{p}})=e^{-\beta {
\hat{\sf{p}}^2
\over
     2M
}
}.
\end{equation}
Further formal properties of (\ref{eq}) are discussed in~\cite{art5,art8}.
\par
\section{FOKKER-PLANCK EQUATION FOR THE \\ DESCRIPTION OF QUANTUM
  DISSIPATION}
\setcounter{equation}{0}
\par
Given the master-equation (\ref{superprl}) one is naturally led to the
question, whether some small parameter having a definite physical
meaning exists, allowing for a Kramers-Moyal expansion leading from
the master-equation to a Fokker-Planck equation~\cite{vanKampen}. This
is in fact the case for the momentum transfer ${\bf q}$, small ${\bf
  q}$ corresponding through the physical meaning of the dynamic
structure factor to the long wavelength part of the density
fluctuations' spectrum of the macroscopic system with which the
Brownian particle is interacting. In the limit of small momentum
transfer, keeping terms at most second order as typical in
Fokker-Planck equations~\cite{Risken}, the operator part of
(\ref{superprl}) becomes
\begin{eqnarray*}
&&
\!\!\!\!\!\!\!\!\!\!\!\!\!\!\!\!\!\!\!\!\!\!\!\!
  \Biggl[
        \exp\left({{\cal L}^{\scriptscriptstyle -}_{\scriptscriptstyle
              {{\bf q}}\cdot{\hat {{\sf x}}}}}\right) 
        \exp\left({{\cal L}^{\scriptscriptstyle +}_{\scriptscriptstyle
              \kappa{{\bf q}}\cdot{\hat {{\sf p}}}}}\right) 
        [{\hat \varrho}]
        -
        \frac 12
        \left \{
      \exp\left({\frac{2\kappa}{\hbar}{{\bf q}}\cdot{\hat {{\sf p}}}}\right)
          ,
      {\hat \varrho}
        \right \}
        \Biggr]\approx
\\
        &\approx&
        {{\cal L}^{\scriptscriptstyle -}_{\scriptscriptstyle {{\bf
                q}}\cdot{\hat {{\sf x}}}}}[{\hat \varrho}]+ 
        \frac{1}{2}{{\cal L}^{\scriptscriptstyle
            - \scriptstyle 2}_{\scriptscriptstyle {{\bf q}}\cdot{\hat {{\sf
                  x}}}}}[{\hat \varrho}]+ 
        {{\cal L}^{\scriptscriptstyle +}_{\scriptscriptstyle
            \kappa{{\bf q}}\cdot{\hat {{\sf p}}}}}[{\hat \varrho}]+ 
        \frac{1}{2}{{\cal L}^{\scriptscriptstyle
            +\scriptstyle 2}_{\scriptscriptstyle \kappa{{\bf
                q}}\cdot{\hat {{\sf 
                  p}}}}}[{\hat \varrho}]
        + 
        {{\cal L}^{\scriptscriptstyle -}_{\scriptscriptstyle {{\bf
                q}}\cdot{\hat {{\sf x}}}}}{{\cal
            L}^{\scriptscriptstyle +}_{\scriptscriptstyle \kappa{{\bf
                q}}\cdot{\hat {{\sf p}}}}}[{\hat \varrho}] 
        - {{\cal L}^{\scriptscriptstyle +}_{\scriptscriptstyle
            \kappa{{\bf q}}\cdot{\hat {{\sf p}}}}}[{\hat \varrho}]
        -{{\cal L}^{\scriptscriptstyle +}_{\scriptscriptstyle
           {\kappa^2/ \hbar} ({{\bf q}}\cdot{\hat {{\sf
                 p}}})^2}}[{\hat \varrho}]
       \\
       &=&
        \sum_{i=1}^3 {{\bf q}}_i {{\cal L}^{\scriptscriptstyle
            -}_{\scriptscriptstyle\hat{{\sf x}}_i }}[{\hat \varrho}] 
        +\frac{1}{2}
         \sum_{i,j=1}^3 {{\bf q}}_i {{\bf q}}_j
{         \left\{
{{\cal L}^{\scriptscriptstyle
            -}_{\scriptscriptstyle\hat{{\sf x}}_i }}{{\cal
            L}^{\scriptscriptstyle 
            -}_{\scriptscriptstyle\hat{{\sf x}}_j }}[{\hat \varrho}] 
        +{{\cal L}^{\scriptscriptstyle
            -}_{\scriptscriptstyle\kappa\hat{{\sf p}}_i }}{{\cal
            L}^{\scriptscriptstyle 
            -}_{\scriptscriptstyle\kappa\hat{{\sf p}}_j }}[{\hat \varrho}] 
        +{{\cal L}^{\scriptscriptstyle
            -}_{\scriptscriptstyle\hat{{\sf x}}_i }}{{\cal
            L}^{\scriptscriptstyle 
            +}_{\scriptscriptstyle 2\kappa\hat{{\sf p}}_j }}[{\hat
          \varrho}]  
\right\}     }.        
\end{eqnarray*}
Integrating over ${{\bf q}}$ only terms bilinear in the momentum
transfer with $i=j$ survive, and exploiting further the isotropy of
the gas implying  ${{\bf q}}^2_i = \frac 13 q^2 $ one obtains
\begin{eqnarray}
        \label{fpl}
        {  
        d {\hat \varrho}  
        \over  
                      dt
        }  
        &=&
        {}-
        {i \over \hbar}
        \left[{\hat {{\sf H}}}_0
        ,
        {\hat \varrho}
        \right]
        +
        z\frac{2}{3}{\pi^2 m^2 \over\beta\hbar}
        \int_{{\bf R}^3} d^3\!
        {{\bf q}}
        \,  
        {
        | \tilde{t} (q) |^2
        q
        }
        e^{-
        {
        \beta
        \over
             8m
        }
        {(1+2\alpha) {{q}}^2}
        }
        \\
        &&
        \hphantom{cosicosi}
        \times
        \sum_{i=1}^3{         \left\{
{{\cal L}^{\scriptscriptstyle
            -\scriptstyle 2}_{\scriptscriptstyle\hat{{\sf x}}_i
            }}[{\hat \varrho}]  
        +{{\cal L}^{\scriptscriptstyle
            -\scriptstyle 2}_{\scriptscriptstyle\kappa\hat{{\sf p}}_i
            }}[{\hat \varrho}]  
        +{{\cal L}^{\scriptscriptstyle
            -}_{\scriptscriptstyle\hat{{\sf x}}_i }}{{\cal
            L}^{\scriptscriptstyle 
            +}_{\scriptscriptstyle 2\kappa\hat{{\sf p}}_i }}[{\hat
          \varrho}]  
\right\}     }.
        \nonumber
\end{eqnarray}

We now want to evaluate the overall coefficient for some cases of
physical interest. Before this we note that the linear dependence on
the fugacity $z$ in (\ref{fpl}) is a result typical of a gas of
Maxwell-Boltzmann particles. Keeping effects due to quantum statistics 
into account~\cite{art4} the factor $z$ has to be replaced by a
function $\zeta (z)$ defined in the following way
\begin{displaymath}
    \zeta (z)=
   \left\{ \matrix{ z & \qquad {\rm Maxwell-Boltzmann}  \cr 
       z/ (1-z) & {\rm Bose} \cr z/ (1+z)  & {\rm Fermi} }\right.
,
\end{displaymath}
so that we will generally consider the coefficient
\begin{equation}
  \label{dpp}
          D_{pp}
        =\zeta (z)
        \frac 23
        {\pi^2 m^2 \over\beta\hbar}
        \int_{{\bf R}^3} d^3\!
        {\bf{q}}
        \,  
        {
        | \tilde{t} (q) |^2
        }
        q
        e^{-
        {
        \beta
        \over
             8m
        }
        {{{q}}^2}
        }.
\end{equation}
We will give two examples. We consider first the case of a short range 
potential characterized by a strength $v_0$ and a typical range $r_0$, 
according to
\begin{equation}
  \label{p1}
  t (\textbf{x})=v_0 e^{-{|\textbf{x}|^2}/{r_0^2}}.
\end{equation}
The Fourier transform of (\ref{p1}) is given by
\begin{displaymath}
   \tilde{t} (\textbf{q})=        \int_{{\bf R}^3} 
        d^3\!
        {\bf{x}}
        \,
        {
        e^{{i\over\hbar}{\bf{q}}\cdot{{\bf{x}}}}  
        \over  
        (2\pi\hbar)^3  
        }
        t (\textbf{x})=
        \frac{\pi^{3/2}}{(2\pi\hbar)^3}v_0
        r_0^3e^{-\frac{q^2r_0^2}{4\hbar^2}} 
\end{displaymath}
and the coefficient (\ref{dpp}) becomes accordingly
\begin{displaymath}
D_{pp}
        =\zeta (z)\frac{1}{48}  v_0^2
        \frac{m}{\hbar}\frac{\upsilon^3}{(1+\upsilon)^2} 
\end{displaymath}
with $\upsilon$ a characteristic constant given by the square ratio
between potential range and thermal wavelength
$\lambda_{\scriptscriptstyle T}=\sqrt{2\pi \beta\hbar^2/m}$ of the
particles of the gas
\begin{equation}
  \label{upsilon}
  \upsilon=8\pi\frac{r_0^2}{\lambda^2_{\scriptscriptstyle T}}.
\end{equation}
As a second example we consider the case in which the range of the
potential shrinks to zero, so that the collisions are described by an
effective T-matrix of the form
\begin{equation}
  \label{p2}
  t (\textbf{x})=\frac{2\pi\hbar^2}{M}a_0\delta^3 (\textbf{x}),
\end{equation}
where $a_0$ is a characteristic scattering length. The Fourier
transform of (\ref{p2}) is 
\begin{displaymath}
   \tilde{t} (\textbf{q})= \frac{1}{4\pi^2}\frac{a_0}{\hbar M}
\end{displaymath}
and as a consequence
\begin{displaymath}
D_{pp}
        =\zeta (z)\frac{32}{3}
        \frac{m}{\hbar\beta^2}\alpha^2\frac{a_0^2}
        {\lambda^2_{\scriptscriptstyle T}}.  
\end{displaymath}
As it can be seen, given some exact expression or some
phenomenological Ansatz for the T-matrix describing the collisions,
one obtains a definite expression for the coefficient $D_{pp}$, which
as we shall see is connected to diffusion in momentum, depending on
the physical parameters of interest.

To make the comparison with the literature easier (\ref{fpl}) using
(\ref{super}) can be also written
\begin{eqnarray}
        \label{proalberto}
        {  
        d {\hat \varrho}  
        \over  
                dt  
        }  
        =
        &-&
        {i\over\hbar}
        [
        {{\hat {{\sf H}}}_0}
        ,{\hat \varrho}
        ]
        \nonumber \\
        &-&
        D_{pp}\sum_{i=1}^3\left\{\frac{1}{\hbar^2}        \left[  
        {\hat {{\sf x}}}_i,
        \left[  
        {\hat {{\sf x}}}_i,{\hat \varrho}
        \right]  
        \right]  
        +\frac{\kappa^2}{\hbar^2}         \left[  
        {\hat {\sf p}}_i,
        \left[  
        {\hat {\sf p}}_i,{\hat \varrho}
        \right]  
        \right]  
        -
        {i\over\hbar^2}2\kappa
        \left[  
        {\hat {{\sf x}}}_i ,
        \left \{  
        {\hat {\sf p}}_i,{\hat \varrho}
        \right \}  
        \right]
        \right\}
      .
\end{eqnarray}
The Fokker-Planck equation (\ref{proalberto}) gives an example of
unbounded generator of a completely positive
quantum dynamical semigroup and corresponds to a particular physical
realization of the diffusive continuous component of the general
structure of translation-covariant quantum dynamical semigroup
characterized by Holevo~\cite{HolevoTI}. In fact (\ref{proalberto}) is
invariant under  
translations according to (\ref{trasla}), moreover an operator of the
form (\ref{equi}) is still a stationary solution due to the particular 
ratio between the friction coefficient and the coefficient of
diffusion in momentum~\cite{Tannor97}, as discussed in the
following. To draw a 
connection with the classical description of Brownian motion the last
three terms of  (\ref{proalberto}) can be recognized as being due to
diffusion in momentum, diffusion in position and friction
respectively~\cite{garda00}. In fact exploiting the correspondence
principle the commutator with the position operator corresponds to a
derivative with respect to momentum, the commutator with the momentum
operator corresponds to a 
derivative with respect to position and the anti-commutator with the
momentum operator corresponds to a linear multiplication by momentum
with a factor two, as can also be most directly seen in terms of the
Wigner function~\cite{Isar99,art6-art7}. In particular the ratio between
the coefficient responsible for diffusion in momentum and the
coefficient responsible for friction is given by $M/\beta$ as in the
classical Kramers' equation for Brownian motion in
phase space~\cite{Kramers,vanKampen}, thus granting the expected
stationary solution (\ref{equi}).

Eq.~(\ref{proalberto}) is a particular realization, obtained on the
basis of a microphysical model, of the general phenomenological
expression for quantum Brownian motion considered
in~\cite{AlbertoQBM} as a starting point for the application of the
theory of measurement continuous in time. It does provide a physically 
motivated structure of generator of quantum dynamical semigroup
allowing for the introduction of an objective description in terms of
trajectories in the sense clarified in~\cite{Lanz94}.

\par
\section{ACKNOWLEDGEMENTS}
\par
The author would like to thank Prof. L. Lanz for useful discussions on
the subject of this paper. The work was supported by MURST under
Cofinanziamento and 
Progetto Giovani.

\par

\end{document}